\documentclass[aps,prl,twocolumn,superscriptaddress,nofootinbib]{revtex4-1}
\usepackage{mathrsfs}
\usepackage{amsfonts}
\usepackage{amsmath}
\usepackage{txfonts}
\usepackage{amssymb}
\usepackage{graphicx}
\usepackage{bm}
\usepackage{color}

\newcommand{\ket}[1]{|#1\rangle}
\newcommand{\bra}[1]{\langle #1|}

\begin{document}
\title{Environment-assisted holonomic quantum maps}
\author{Nicklas Ramberg}
\affiliation{Department of Physics and Astronomy, Uppsala University, Box 516,
Se-751 20 Uppsala, Sweden}
\author{Erik Sj\"{o}qvist}
\email{erik.sjoqvist@physics.uu.se}
\affiliation{Department of Physics and Astronomy, Uppsala University, Box 516,
Se-751 20 Uppsala, Sweden}
\date{\today}
\begin{abstract} 
Holonomic quantum computation uses non-Abelian geometric phases to realize error resilient 
quantum gates. Nonadiabatic holonomic gates are particularly suitable to avoid unwanted 
decoherence effects, as they can be performed at high speed. By letting the computational 
system interact with a structured environment, we show that the scope of error resilience of 
nonadiabatic holonomic gates can be widened to include systematic parameter errors. Our 
scheme maintains the geometric properties of the evolution and results in an  
environment-assisted holonomic quantum map that can mimic the effect of a holonomic 
gate. We demonstrate that the sensitivity to systematic errors 
can be reduced in a proof-of-concept spin-bath model.    
\end{abstract}
\maketitle
\date{\today}
Quantum holonomies are non-Abelian (non-commuting) unitary operators  that only depend 
on paths in state space of a quantum system. The non-commuting property makes them useful 
for implementing quantum gates that manipulate quantum information by purely geometric means. 
Holonomic quantum computation (HQC) \cite{zanardi99} is a network of holonomic gates that 
unifies geometric characteristics of quantum systems and information processing, as 
well as is conjectured to be robust to errors in experimental control parameters \cite{pachos01}. 

Nonadiabatic HQC has recently been proposed \cite{sjoqvist12} and experimentally 
implemented \cite{abdumalikov13,feng13,arroyo14,zu14,danilin18,xu18a} as a tool to 
realize quantum gates based upon nonadiabatic non-Abelian geometric phases 
\cite{anandan88}. The basic setup for nonadiabatic HQC in \cite{sjoqvist12} is a three-level 
$\Lambda$ configuration, where two simultaneous resonant laser pulses 
drive transitions between the qubit levels and an auxiliary state level. This scheme has 
been generalized to off-resonant pulses \cite{xu15,sjoqvist16}. The off-resonant 
setup uses two simultaneous laser pulses with the same variable detuning, 
which enhances the flexibility of the holonomic scheme. For experimental realization of 
off-resonant nonadiabatic holonomic gates, see Refs.~\cite{zhou17,sekiguchi17,li17,zhang18}. 

The nonadiabatic version of HQC avoids the drawback of the long run time associated with 
adiabatic holonomies \cite{wilczek84}, on which the original holonomic schemes are based 
\cite{zanardi99,duan01}. Nonadiabatic holonomic gates are therefore particularly suitable to 
avoid unwanted decoherence effects \cite{johansson12}. The resilience to decoherence errors 
can be further improved by combining nonadiabatic HQC with decoherence-free subspaces 
\cite{xu12,liang14,xue15,zhao17} and subsystems \cite{zhang14}, as well as  dynamical 
decoupling \cite{xu14,sun16,xu18b}. On the other hand, it has been pointed out \cite{zheng16} 
that the original version of nonadiabatic HQC has no particular advantage 
compared to standard dynamical schemes in the presence of systematic errors in experimental 
parameters. To deal with this, we here show that the sensitivity to systematic parameter errors 
can be reduced by letting the system interact with a structured environment. Our approach is 
inspired by earlier findings \cite{sinayskiy12,giorgi12,marais13} that transport efficiency in 
quantum systems can be enhanced in such environments.  

We modify the off-resonant non-adiabatic holonomic scheme by coupling the auxiliary state 
to a finite thermal bath, the latter playing the role of the structured environment. The key point 
of our modified scheme is its non-Markovian nature. This allows for coherence to flow back 
and forth between the system and the environmental bath, a feature that has been shown to 
prolong coherence \cite{jing10,jing12}, and can be used for quantum control \cite{luo15,luo18} 
in open quantum systems. The resulting transformation retains 
its holonomic property  and can therefore be regarded as an environment-assisted holonomic 
map that can mimic the effect of a holonomic gate. We address the protective potential of 
the environment to errors in the Rabi frequencies and detuning describing the system-laser 
interaction, which causes the qubits to end up only partially in the computational subspace, 
while preserving the purely geometric property of the evolution. Our aim is to demonstrate 
that the sensitivity to systematic deviations in the Rabi frequencies and detuning can be 
reduced by tuning the system-bath coupling strength to its optimal value. 

We consider an off-resonant $\Lambda$ system, in which two square-shaped  simultaneous 
laser pulses induce transitions between the computational state levels $\ket{j}$, $j=0,1$, and 
an auxiliary state $\ket{e}$. The corresponding Rabi frequencies take the form $\Omega_0 = \omega 
e^{i\varphi} \sin \frac{\theta}{2}$ and $\Omega_1 = -\omega \cos \frac{\theta}{2}$ with $\theta$ 
and $\varphi$ controlling the relative amplitude and phase, respectively, of the two pulses. 
The Hamiltonian during the pulse pair reads ($\hbar = 1$ from now on)
\begin{eqnarray}
H_{\Lambda} = \delta \ket{e} \bra{e} + \omega \big( \ket{e} \bra{b} + \ket{b} \bra{e} \big) , 
\end{eqnarray}
where $\ket{b} = e^{-i\varphi} \sin \frac{\theta}{2} \ket{0} - \cos \frac{\theta}{2} \ket{1}$ 
is the bright state, while the dark state $\ket{d} = \cos \frac{\theta}{2} \ket{0} + 
e^{i\varphi} \sin \frac{\theta}{2} \ket{1}$ is decoupled from the evolution. We assume the same 
detuning $\delta$ of the two transitions. Provided the parameters $\delta,\omega,\theta$, and 
$\varphi$ are kept constant during the pulse pair, the resulting evolution is purely geometric as the 
Hamiltonian vanishes on the evolving computational subspace
$\mathcal{M}_c (t) = {\rm Span} \{ U_{\Lambda}(t,0) \ket{0}, U_{\Lambda}(t,0) \ket{1} \}$, 
$U_{\Lambda}(t,0)$ being the 
time evolution operator associated with $H_{\Lambda}$. Ideally, the pulse duration $\tau_0$ 
should be chosen so as to ensure cyclic evolution, i.e., $\bra{e} U(\tau_0,0) \ket{b} = 0$, 
which yields $\tau_0 = 2\pi/\Delta_0$ with $\Delta_0 = \sqrt{\delta^2 + 4\omega^2}$. This  
results in the holonomic gate \cite{xu15,sjoqvist16}
\begin{eqnarray}
U(C) = \ket{d} \bra{d} - e^{-i\chi} \ket{b} \bra{b}  = 
e^{i\frac{1}{2} (\pi - \chi)} e^{-i\frac{1}{2} (\pi - \chi) {\bf n} \cdot \boldsymbol{\sigma}} , 
\label{eq:idealgate} 
\end{eqnarray}
which acts on $\mathcal{M}_c (\tau_0) = \mathcal{M}_c (0)$. Here, ${\bf n} = 
(\sin \theta \cos \varphi , \sin \theta \sin \varphi , \cos \theta)$ is the gate 
rotation axis and $C$ is the path traced out by $\mathcal{M}_c (t)$ in the Grassmannian 
$\mathcal{G} (3;2)$, i.e., the space of two-dimensional subspaces of the three-dimensional 
space ${\rm Span} \{ \ket{0},\ket{1},\ket{e}\}$. The phase $\chi = (\delta/2) \tau_0$ 
determines the gate rotation angle $\pi - \chi$ around ${\bf n}$. 

Next, we introduce the environmental bath $B$ with Hilbert space $\mathcal{H}_B$, 
assuming $\dim \mathcal{H}_B = K$ finite. The Hamiltonian during the pulse pair is assumed 
to take the form 
\begin{eqnarray}
H = H_{\Lambda} \otimes \hat{1}_B + \hat{1}_{\Lambda} \otimes H_B + 
\gamma \ket{e} \bra{e} \otimes h_B  
\label{eq:hamiltonian}
\end{eqnarray}
with $H_B$ and $h_B$ both time-independent, and $\gamma$ the system-bath coupling 
strength. The spectrum of $h_B$ is  $\mu_0 = 0 \leq \mu_1 \leq \ldots \leq \mu_{K-1}$  
with corresponding eigenstates $\ket{\mu_0},\ket{\mu_1},\ldots,\ket{\mu_{K-1}}$. We 
may rewrite the Hamiltonian during the pulse as  
\begin{eqnarray}
H = \sum_{k=0}^{K-1} H_{\delta + \gamma\mu_k} (\omega,\theta,\varphi) \otimes 
\ket{\mu_k} \bra{\mu_k} + \hat{1}_{\Lambda} \otimes H_B , 
\end{eqnarray}
where 
\begin{eqnarray}
H_{\delta + \gamma \mu_k} (\omega,\theta,\varphi) = (\delta + \gamma \mu_k) \ket{e} \bra{e} + 
\omega (\ket{e} \bra{b} + \ket{b} \bra{e}).
\end{eqnarray}
Each sub-Hamiltonian $H_{\delta + \gamma \mu_k} (\omega,\theta,\varphi)$ induces a 
purely geometric evolution along a path $C_k$ in $\mathcal{G} (3;2)$. 
The corresponding holonomic sub-gate $U(C_k)$ is obtained by replacing $\delta$ with 
$\delta + \gamma\mu_k$ in the above ideal gate Eq.~(\ref{eq:idealgate}). Note that $C_k$ 
are paths  associated with different cyclic times, due to the $\mu_k$-dependence of the 
modified detunings $\delta + \gamma\mu_k$. 

We assume the bath starts in a thermal state $\rho_{\beta}$ that factorizes with the initial 
pure system state $\ket{\psi}$ in $\mathcal{M}_c (0)$. In other words, the full system-bath 
state $\ket{\psi}\bra{\psi} \otimes \rho_{\beta}$ evolves as 
\begin{eqnarray}
\varrho (t) = U(t,0) \big( \ket{\psi} \bra{\psi} \otimes \rho_{\beta} \big) U^{\dagger} (t,0)  
\end{eqnarray}
with 
\begin{eqnarray}
\varrho_{\beta} = \frac{1}{Z} e^{-\beta H_B} ,  
\end{eqnarray}
$Z = {\rm Tr} \left( e^{-\beta H_B} \right)$ being the partition function and $\beta^{-1}$ 
the temperature. $U(t,0) = e^{-iHt}$ is the time evolution operator with $H$ given by 
Eq.~(\ref{eq:hamiltonian}). The computational input state evolves as 
\begin{eqnarray}
\ket{\psi} \bra{\psi} \mapsto \rho (t) = {\rm Tr}_B \varrho (t) ,
\end{eqnarray}
where ${\rm Tr}_B$ is partial trace over the bath. In the case where $[H_B,h_B] = 0$, 
we may explicitly evaluate the partial trace yielding the computational state 
\begin{eqnarray}
\rho (t) = \sum_{k = 0}^{K-1} \frac{e^{-\beta \nu_k}}{Z} 
e^{-iH_{\delta + \gamma\mu_k} (\omega,\theta,\varphi) t} \ket{\psi} \bra{\psi} 
e^{iH_{\delta + \gamma\mu_k} (\omega,\theta,\varphi) t} , 
\end{eqnarray} 
where we assumed the spectrum $\nu_0 , \nu_1 , \ldots , \nu_{K-1}$ of $H_B$. 
This is a unital map $\rho \mapsto \mathcal{E}_t (\rho)$ with Kraus operators 
\begin{eqnarray}
A_k (t) = \sqrt{\frac{e^{-\beta \nu_k}}{Z}} 
e^{-iH_{\delta + \gamma\mu_k} (\omega,\theta,\varphi) t} .
\end{eqnarray}
The map $\mathcal{E}_t$ is the promised environment-assisted holonomic quantum map.  

For very low temperatures, only the ground state of the bath is populated and the 
system undergoes unitary evolution governed by the Hamiltonian $H_{\Lambda}$. The 
resulting gate is essentially determined by the Rabi frequencies $\Omega_j$ and the 
detuning $\delta$. These parameters can be affected by errors: 
\begin{eqnarray}
\Omega'_j =  (1+\epsilon_j) e^{i\zeta_j} \Omega_j , \ \ \delta' = (1+\kappa) \delta , 
\end{eqnarray}
$\epsilon_j,\zeta_j$, and $\kappa$ being real-valued numbers. This can be translated into the 
parameters 
\begin{eqnarray}
\omega' & = & \left[ (1+\epsilon_0)^2 \sin^2 \frac{\theta}{2} + 
(1+\epsilon_1)^2 \cos^2 \frac{\theta}{2} \right]^{1/2} \omega, 
\nonumber \\ 
e^{i\varphi'} \tan \frac{\theta'}{2} & = & \left( \frac{1+\epsilon_0}{1+\epsilon_1} \right) 
e^{i(\zeta_0-\zeta_1)} e^{i\varphi} \tan \frac{\theta}{2} . 
\end{eqnarray}
The dark and bright statets are modified accordingly, i.e., they read $\ket{d'} = 
\cos \frac{\theta'}{2} \ket{0} + e^{i\varphi'} \sin \frac{\theta'}{2} \ket{1}$ and $\ket{b'} = 
e^{-i\varphi'} \sin \frac{\theta'}{2} \ket{0} - \cos \frac{\theta'}{2} \ket{1}$. 

We assume $\epsilon_j,\zeta_j$, and $\kappa$ are constant, which correspond to systematic 
errors in the applied pulse pair. Under this assumption, the evolution remains purely 
geometric since the error-affected Hamiltonian $H_{\delta'} (\omega',\theta',\varphi')$ 
remains zero on $\mathcal{M}_c (t)$. The errors in $\theta$ and $\varphi$ change the direction 
${\bf n}$ of the rotation axis, but preserve the cyclic property of the evolution. On the other 
hand, the computational subspace would generally fail to return at $t=\tau_0$ due to 
errors in $\omega$ and $\delta$. 

We shall investigate the performance of the environment-assisted holonomic maps in the 
presence of Rabi frequency errors, by comparing them to the ideal gate $U(C)$ with run time 
$\tau_0$ by means of the fidelity 
\begin{eqnarray}
\mathcal{F}(\psi) & = & \bra{\psi} U^{\dagger} (C) \rho (\tau_0) U (C) \ket{\psi}^{1/2} 
\nonumber \\ 
 & = &  \left( \sum_{k = 0}^{K-1} \left| \bra{\psi} U^{\dagger} (C) A'_k (\tau_0)
\ket{\psi} \right|^2 \right)^{1/2} 
\label{eq:fidelity}
\end{eqnarray}
with 
\begin{eqnarray}
A'_k (\tau_0) = \sqrt{\frac{e^{-\beta \nu_k}}{Z}} 
e^{-iH_{\delta' + \mu_k} (\omega',\theta',\varphi') \tau_0} . 
\end{eqnarray}
The fidelity $\mathcal{F}(\psi)$ can be studied as a function of coupling parameter $\gamma$, 
for some suitably chosen input states $\ket{\psi}$, given a number of degrees of freedom $N$ 
in the bath and temperature $\beta^{-1}$. As a measure of gate performance, one averages 
the fidelity over a sufficiently large, uniformly distributed sample of input states. 

\begin{figure*}[htb]
\centering
\vspace{-50pt}
\hspace{-100pt} 
\includegraphics[width=0.6\textwidth]{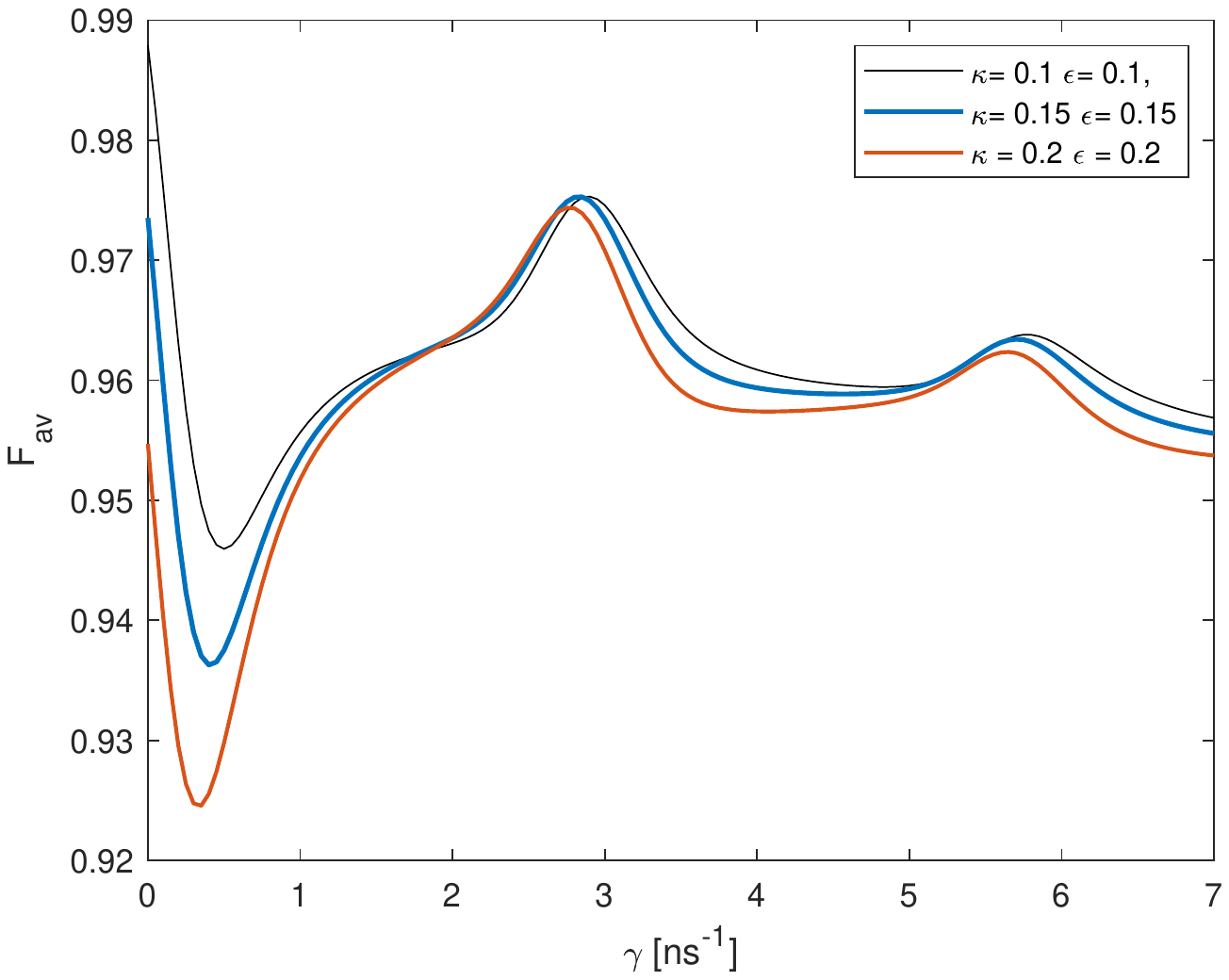}
\hspace{-80pt} 
\includegraphics[width=0.6\textwidth]{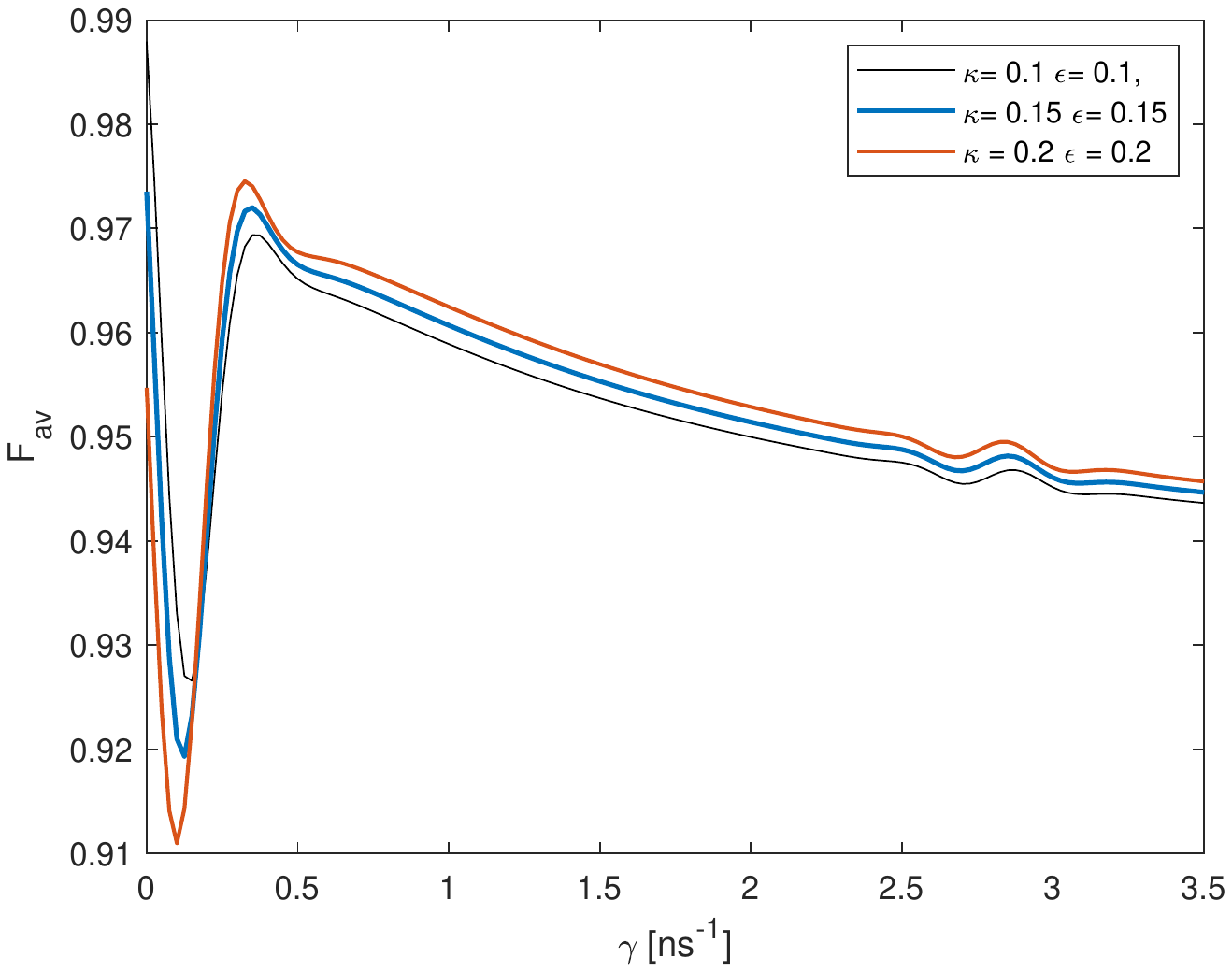}
\vspace{-120pt}
\hspace{-100pt} 
\caption{Gate performance of a spin-bath-based environment-assisted holonomic map 
relative a nonadiabatic holonomic gate with Rabi frequency $\omega = 1 \ {\rm ns}^{-1}$ 
and detuning $\delta = 2 \ {\rm ns}^{-1}$. Average fidelity as a function of system-bath 
coupling with $N=20$ spins and the bath parameter $\alpha = 15 {\rm ps}^{-1}$ at $T=50 \ {\rm K}$ 
(left panel) and $T=300 \ {\rm K}$ (right panel) are shown. The error parameters $\epsilon$ 
and $\kappa$ relate the error affected and ideal Rabi frequency and detuning according to 
$\omega' = (1+\epsilon) \omega$ and $\delta' = (1+\kappa) \delta$, respectively.}
\label{fig:1}
\end{figure*}

To address the behavior of the environment-assisted holonomic scheme in an explicit 
proof-of-concept system, we consider the spin-bath model proposed in Ref.~\cite{sinayskiy12} 
adapted to the $\Lambda$ system. We choose 
\begin{eqnarray}
H_B = \alpha S_z , \ \ 
h_B = \left( S_z + \frac{N}{2} \right) 
\end{eqnarray}
with $S_z$ the $z$-projection of the total spin of the bath consisting of $N$ individual 
spin$-\frac{1}{2}$. The spectrum of $H_B$ is $\nu_m = \alpha (m-N/2)$, $m=0,\ldots,N$, 
with multiplicities $\binom{N}{m}$. In other words, the bath parameter $\alpha$ measures 
the energy split between the eigenstates of $H_B$. The initial thermal bath state reads 
\begin{eqnarray}
\varrho_{\beta} & = &  
\sum_{m=0}^{N} \sum_{q=1}^{\binom{N}{m}} \frac{e^{-\beta \alpha m}}{Z} 
\ket{m-N/2,q} \bra{m-N/2,q} , 
\nonumber \\ 
Z & = & \sum_{m=0}^{N} \binom{N}{m} e^{-\beta \alpha m} ,
\end{eqnarray}
where $\ket{m-N/2,q}$ are eigenstates of $S_z$ consisting of permutations of $m$ 
spins in $\ket{\! \uparrow_z}$ and $N-m$ spins in $\ket{\! \downarrow_z}$. 
The spectrum of $h_B$ is $m$. This defines the error affected unital map 
\begin{eqnarray}
\ket{\psi} \bra{\psi} \mapsto 
\rho (\tau_0) = \sum_{m=0}^N A'_m (\tau_0) \ket{\psi} \bra{\psi} A_m^{' \dagger} (\tau_0)
\end{eqnarray}
with Kraus operators 
\begin{eqnarray}
A'_m (\tau_0) & = & \sqrt{\binom{N}{m} \frac{e^{-\beta\alpha m}}{Z}}  
e^{-iH_{\delta'+\gamma m} (\omega'\theta',\varphi') \tau_0} ,
\nonumber \\ 
H_{\delta'+\gamma m} (\omega',\theta',\varphi') & = & (\delta' + \gamma m) \ket{e} \bra{e} 
\nonumber \\ 
 & & + \omega' (\ket{e} \bra{b'} + \ket{b'} \bra{e}) . 
\end{eqnarray}

To simplify the analysis, we assume $\epsilon_0 = \epsilon_1 \equiv \epsilon$ and 
$\zeta_0- \zeta_1 = 0$, which imply  $\omega' = (1+\epsilon) \omega, \theta' = \theta$, and 
$\varphi' = \varphi$. Under these restrictions, $\ket{d'} = \ket{d}$ and $\ket{b'} = \ket{b}$, 
which imply that the gate rotation axis ${\bf n}'$ coincides with the ideal ${\bf n}$, while 
$\mathcal{M}_c$ undergoes cyclic evolution for the pulse duration $\tau_0'$ being generally 
different from the ideal run time $\tau_0$. In other words, the computational subspace typically 
fails to return after applying the error affected pulse pair for the ideal duration $\tau_0$ at 
zero temperature. 

We now wish to optimize the gate performance by maximizing the similarity between 
the error affected environment-assisted holonomic map with Kraus operators $A'_m (\tau_0)$ 
and the ideal holonomic gate $U(C) = \ket{d} \bra{d} - e^{-i\chi} \ket{b} \bra{b}$ by tuning 
the system-bath coupling strength $\gamma$ at nonzero temperature. To formalize this 
idea, we write $\ket{\psi} = \cos \frac{\vartheta}{2} \ket{d} + e^{i\xi} \sin \frac{\vartheta}{2} \ket{b}$ 
and obtain  
\begin{eqnarray}
 & & \left| \bra{\psi} U^{\dagger} (C) A'_m (\tau_0) \ket{\psi} \right|^2 = 
 \binom{N}{m} \frac{e^{-\beta \alpha m}}{Z} \left|  \cos^2 \frac{\vartheta}{2} \right. 
\nonumber \\ 
 & & \left. - e^{i\chi} \sin^2 \frac{\vartheta}{2} \bra{b} e^{-iH_{\delta'+\gamma m} 
(\omega', \theta, \varphi) \tau_0} \ket{b} \right|^2 ,
\end{eqnarray}
where we have used that $\bra{d} A'_m (\tau_0) \ket{b} = 0$. We find 
\begin{eqnarray}
\bra{b} e^{-iH_{\delta'+\gamma m} (\omega', \theta, \varphi) \tau_0} \ket{b} & = &  
e^{-i\Sigma'_{\gamma m} \tau_0} \left( \cos \frac{ \pi \Delta'_{\gamma m}}{\Delta_0} \right.  
\nonumber \\ 
 & & \left. + i \cos \eta'_{\gamma m} \sin \frac{\pi \Delta'_{\gamma m}}{\Delta_0} \right) , 
\end{eqnarray}
where the parameters  
\begin{eqnarray}
\tan \eta'_{\gamma m} & = & \frac{2\omega'}{\delta'+\gamma m} , \ \Sigma'_{\gamma m} = 
\frac{\delta' + \gamma m}{2},   
\nonumber \\ 
\Delta'_{\gamma m} & = & \sqrt{ (\delta' + \gamma m)^2 + 4(\omega')^2} 
\end{eqnarray} 
are associated with the diagonalization of $H_{\delta'+\gamma m} (\omega', \theta, \varphi)$. 
The fidelity $\mathcal{F} (\psi) \equiv \mathcal{F} (\vartheta)$ is independent of $\xi$, which 
implies that we only need to sample over $\vartheta$. A uniform distribution of states $\psi$ 
would correspond to a weight factor that is proportional to the circumference of the 
circle at this latitude on the Bloch sphere, i.e., we may take $w(\vartheta) = \sin \vartheta$. 
By  choosing $n$ input states at $\vartheta_k = k\pi /(n-1)$, $k=0,\ldots,n-1$, the averaged 
fidelity thus reads  
\begin{eqnarray}
{\rm F}_{{\rm av}} = 
\frac{\sum_{k=0}^{n-1} \sin  \left( \frac{k\pi}{n-1} \right) \mathcal{F}  
\left( \frac{k\pi}{n-1} \right)}{\sum_{k=0}^{n-1} 
\sin \left( \frac{k\pi}{n-1} \right)} . 
\end{eqnarray}
We measure the gate performance in terms of ${\rm F}_{{\rm av}}$. 

Figure \ref{fig:1} shows the average fidelity as a function of system-bath coupling at 
$T=50 \ {\rm K}$ (left panel) and room temperature $T=300 \ {\rm K}$ (right panel) with 
$N=20$ spins and the bath parameter $\alpha = 15 {\rm ps}^{-1}$. Ideal parameter values are 
chosen to be $\omega = 1 \ {\rm ns}^{-1}$ and $\delta = 2 \ {\rm ns}^{-1}$, corresponding to 
the rotation angle $\pi - \chi \approx 0.29 \pi$. Averages are computed for $n = 30$ 
equidistant $\vartheta$ values. 

\begin{figure}[htb]
\centering
\includegraphics[width=0.42\textwidth]{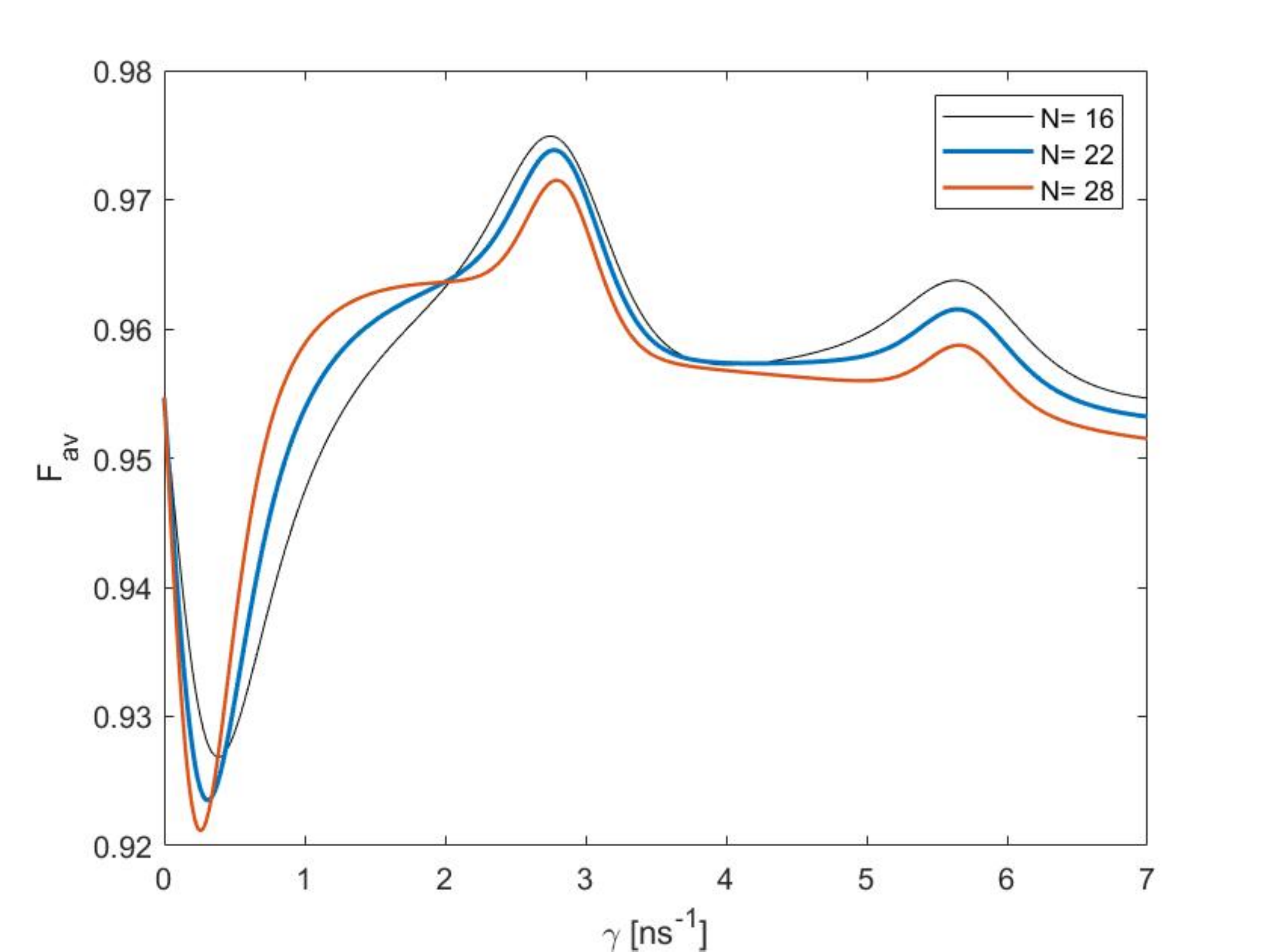}
\caption{Fidelity as a function of system bath coupling $\gamma$ for $N=16,22,28$ spins 
in the bath. The error parameters have been chosen as $\epsilon = \kappa = 0.2$ and 
temperature $T=50 \ {\rm K}$. The optimal coupling strength varies less than $3 \%$ over the 
chosen range of $N$.}
\label{fig:2}
\end{figure}

For both temperatures, we numerically compute the fidelity for errors $\epsilon = \kappa = 
0.1, 0.15,$ and $0.2$. We see that the optimal nonzero system-bath coupling strength  
depends significantly on temperature, but is quite insensitive to the error size. This insensitivity 
holds also for the corresponding optimal fidelity, especially for the lower temperature, 
where ${\rm F}_{\rm av} \sim 97.3 - 97.4 \%$ at the optimal system-bath coupling strength 
$\gamma \sim 2.8 \ {\rm ns}^{-1}$. On the other hand, the fidelity is strongly error-size-dependent 
in the absence of the bath ($\gamma = 0$), in case of which ${\rm F}_{\rm av}$ varies between 
$95.5 \%$ and $98.6 \%$ over the chosen error range. Thus, the environmental bath can 
be made to reduce the sensitivity to systematic errors by tuning  the system-bath coupling 
strength. For large errors ($\epsilon,\kappa \sim 0.15$ or higher) the fidelity takes a higher 
value for the nonzero optimal $\gamma$, which shows that the bath not only can reduce the 
error sensitivity but also can improve the gate performance. This demonstrates that our 
scheme can protect against large systematic errors by fixing the system-bath coupling 
strength at the optimal value for a given bath temperature. Finally, we have tested the 
$N$ dependence of the optimal $\gamma$ value. Figure \ref{fig:2} shows the fidelity 
for $N=16, 22,$ and $28$ in the case where $\epsilon = \kappa = 0.2$ and $T=50 \ {\rm K}$. 
We see that this optimal value varies between $2.74 - 2.80 \ {\rm ns}^{-1}$ and thus depends 
only weakly on $N$. This demonstrates that the exact number of 
spins in the bath seems not essential for the tuning of the system-bath coupling, a feature 
that would simplify the optimization of the environment-assisted holonomic gate.   

In conclusion, we have addressed the sensitivity to systematic parameter errors in nonadiabatic 
holonomic schemes. To this end, we have put forward a concept of environment-assisted 
holonomic maps, in which the auxiliary state of the standard $\Lambda$ system realization 
of holonomic gates is coupled to a finite thermal bath system. These maps retain the geometric 
properties of the ideal holonomic gates. By tuning the system-bath coupling strength to its 
optimal value, the sensitivity to systematic errors can be reduced and the corresponding 
optimal fidelity may in some cases be even higher than in the absence of the environmental 
bath. These features may persist even at room temperature.  We have demonstrated the 
robustness in a proof-of-concept spin-bath model. More sophisticated models, with a larger 
number of optimization parameters and thereby more possible routes toward higher error 
resilience, can be envisaged. 

\section*{Acknowledgments} 
E.S. acknowledges financial support from the Swedish Research Council (VR) through 
Grant No. 2017-03832.


\begin{thebibliography}{99}
\bibitem{zanardi99} P. Zanardi and M. Rasetti, 
Holonomic quantum computation, 
Phys. Lett. A {\bf 264}, 94 (1999).
\bibitem{pachos01} J. Pachos and P. Zanardi, 
Quantum holonomies for quantum computing, 
Int. J. Mod. Phys. B {\bf 15}, 1257 (2001).
\bibitem{sjoqvist12} E. Sj\"oqvist, D. M. Tong, L. M. Andersson, B. Hessmo, 
M. Johansson, and K. Singh,  
Non-adiabatic holonomic quantum computation, 
New J. Phys. {\bf 14}, 103035 (2012).
\bibitem{abdumalikov13} A. A. Abdumalikov, J. M. Fink, K. Juliusson, M. Pechal, S. Berger, 
A. Wallraff, and S. Filipp, 
Experimental realization of non-Abelian non-adiabatic geometric gates, 
Nature (London) {\bf 496}, 482 (2013). 
\bibitem{feng13}  G. Feng, G. F. Xu, and G. L. Long, 
Experimental Realization of Nonadiabatic Holonomic Quantum Computation, 
Phys. Rev. Lett. {\bf 110}, 190501 (2013). 
\bibitem{arroyo14} S. Arroyo-Camejo, A. Lazariev, S. W. Hell, and G. Balasubramanian, 
Room temperature high-fidelity holonomic single-qubit gate on a solid-state spin, 
Nat. Commun. {\bf 5}, 4870 (2014).
\bibitem{zu14} C. Zu, W. B. Wang, L. He, W. G. Zhang, C. Y. Dai, F. Wang, and L. M. Duan, 
Experimental realization of universal geometric quantum gates with solid-state spins, 
Nature (London) {\bf 514}, 72 (2014).
\bibitem{danilin18} S. Danilin, A. Veps\"al\"ainen, and G. S. Paraoanu, 
Experimental state control by fast non- Abelian holonomic gates with a superconducting qutrit, 
Phys. Scr. {\bf 93} 055101 (2018).
\bibitem{xu18a} Y. Xu, W. Cai, Y. Ma, X. Mu, L. Hu, Tao Chen, H. Wang, Y.P. Song, Z.-Y. Xue, 
Z. Yin, and L. Sun, 
Single-Loop Realization of Arbitrary Nonadiabatic Holonomic Single-Qubit Quantum Gates in a Superconducting Circuit,
Phys. Rev. Lett. {\bf 121}, 110501 (2018). 
\bibitem{anandan88} J. Anandan, 
Non-adiabatic non-Abelian geometric phase, 
Phys. Lett. A {\bf 133}, 171  (1988).
\bibitem{xu15} G. F. Xu, C. L. Liu, P. Z. Zhao, and D. M. Tong, 
Nonadiabatic holonomic gates realized by a single-shot implementation, 
Phys. Rev. A {\bf 92}, 052302 (2015).
\bibitem{sjoqvist16} E. Sj\"{o}qvist, 
Nonadiabatic holonomic single-qubit gates in off-resonant $\Lambda$ systems, 
Phys. Lett. A {\bf 380}, 65 (2016).
\bibitem{zhou17} B. B. Zhou, P. C. Jerger, V. O. Shkolnikov, F. J. Heremans, G. Burkard, 
D. D. Awschalom, 
Holonomic Quantum Control by Coherent Optical Excitation in Diamond, 
Phys. Rev. Lett. {\bf 119}, 140503 (2017).
\bibitem{sekiguchi17} Y. Sekiguchi, N. Niikura, R. Kuroiwa, H. Kano, and H. Kosaka, 
Optical holonomic single quantum gates with a geometric spin under a zero field, 
Nature Photonics {\bf 11}, 309 (2017). 
\bibitem{li17} H. Li, Y. Liu, and G. L. Long, 
Experimental realization of single-shot nonadiabatic holonomic gates in nuclear spins, 
Sci. China-Phys. Mech. Astron. {\bf 60}, 080311 (2017). 
\bibitem{zhang18} Z. Zhang, P. Z. Zhao, T. Wang, L. Xiang, Z. Jia, P. Duan, D. M. Tong, 
Y. Yin,  and G. Guo, 
Single-shot realization of nonadiabatic holonomic gates with a superconducting Xmon qutrit, 
arxiv:1811.06252. 
\bibitem{wilczek84} F. Wilczek and A. Zee, 
Appearance of Gauge Structure in Simple Dynamical Systems, 
Phys. Rev. Lett. {\bf 52}, 2111 (1984). 
\bibitem{duan01} L. M. Duan, J. I. Cirac, and P. Zoller, 
Geometric Manipulation of Trapped Ions for Quantum Computation, 
Science {\bf 292}, 1695 (2001).  
\bibitem{johansson12} M. Johansson, E. Sj\"oqvist, L. M. Andersson, M. Ericsson, 
B. Hessmo, K. Singh, and D. M. Tong, 
Robustness of nonadiabatic holonomic gates, 
Phys. Rev. A {\bf 86}, 062322 (2012).
\bibitem{xu12} G. F. Xu, J. Zhang, D. M. Tong, E. Sj\"oqvist, and L. C. Kwek, 
Nonadiabatic Holonomic Quantum Computation in Decoherence-Free Subspaces, 
Phys. Rev. Lett. {\bf 109}, 170501 (2012). 
\bibitem{liang14} Z.-T Liang, Y.-X Du, W. Huang, Z.-Y Xue, and H. Yan, 
Nonadiabatic holonomic quantum computation in decoherence-free subspaces with trapped ions, 
Phys. Rev. A {\bf 89}, 062312 (2014). 
\bibitem{xue15} Z.-Y. Xue, J. Zhou, and Z. D. Wang, 
Universal holonomic quantum gates in decoherence-free subspace on superconducting circuits, 
Phys. Rev. A {\bf 92}, 022320 (2015). 
\bibitem{zhao17} P. Z. Zhao, G. F. Xu, Q. M. Ding, E. Sj\"oqvist, and D. M. Tong, 
Single-shot realization of nonadiabatic holonomic quantum gates in decoherence-free subspaces, 
Phys. Rev. A {\bf 95}, 062310 (2017).
\bibitem{zhang14} J. Zhang, L.-C. Kwek, E. Sj\"oqvist, D. M. Tong, and P. Zanardi, 
Quantum computation in noiseless subsystems with fast non-Abelian holonomies, 
Phys. Rev. A {\bf 89}, 042302 (2014). 
\bibitem{xu14} G. F. Xu and G. L. Long, 
Protecting geometric gates by dynamical decoupling, 
Phys. Rev. A {\bf 90}, 022323 (2014).
\bibitem{sun16} C. Sun, G. Wang, C. Wu, H. Liu, X.-L. Feng, J.-L. Chen, and K. Xue, 
Non-adiabatic holonomic quantum computation in linear system-bath coupling, 
Sci. Rep. {\bf 6}, 20292 (2016). 
\bibitem{xu18b} G. F. Xu, D. M. Tong, and E. Sj\"oqvist, 
Path-shortening realizations of nonadiabatic holonomic gates, 
Phys. Rev. A {\bf 98}, 052315 (2018).
\bibitem{zheng16} S.-B. Zheng, C.-P. Yang, and F. Nori, 
Comparison of the sensitivity to systematic errors between nonadiabatic non-Abelian geometric
gates and their dynamical counterparts, 
Phys. Rev. A {\bf 93}, 032313 (2016). 
\bibitem{sinayskiy12} I. Sinayskiy, A. Marais, F. Petruccione, and A. Ekert, 
Decoherence-Assisted Transport in a Dimer System, 
Phys. Rev. Lett. {\bf 108}, 020602 (2012). 
\bibitem{giorgi12} G. L. Giorgi and T. Busch, 
Decoherence-assisted transport and quantum criticalities, 
Phys. Rev. A {\bf 86}, 052112 (2012). 
\bibitem{marais13} A. Marais, I. Sinayskiy, A. Kay, F. Petruccione, and A. Ekert, 
Decoherence-assisted transport in quantum networks, 
New J. Phys. {\bf 15}, 013038 (2013). 
\bibitem{jing10} J. Jing and T. Yu, 
Non-Markovian Relaxation of a Three-Level System: Quantum Trajectory Approach, 
Phys. Rev. Lett. {\bf 105}, 240403 (2010).  
\bibitem{jing12} J. Jing, X. Zhao, J. Q. You, and T. Yu, 
Time-local quantum-state-diffusion equation for multilevel quantum systems, 
Phys. Rev. A {\bf 85}, 042106 (2012). 
\bibitem{luo15} D.-W. Luo, P. V. Pyshkin, C.-H. Lam, T. Yu, H.-Q. Lin, J. Q. You, and L.-A. Wu, 
Dynamical invariants in a non-Markovian quantum-state-diffusion equation, 
Phys. Rev. A {\bf 92}, 062127 (2015). 
\bibitem{luo18} D.-W. Luo, J. Q. You, H.-Q. Lin, L.-A. Wu, and T. Yu, 
Memory-induced geometric phase in non-Markovian open systems, 
Phys. Rev. A {\bf 98}, 052117 (2018).  
\end{thebibliography}
\end{document}